\documentclass[aps,prl,twocolumn]{revtex4-1}
\usepackage{graphicx}
\usepackage{amsfonts}
\usepackage{amsmath}
\usepackage{color}
\usepackage{bm}
\usepackage{amssymb}

\usepackage{multirow}
\usepackage{epstopdf}
\usepackage{natbib}

\newcommand{\beq}{\begin{equation}}
\newcommand{\eeq}{\end{equation}}

\def\prc#1#2#3{{\it Phys.~Rev.}~C~{\bf #1},\ #2\ (#3)}

\def\jcp#1#2#3{{\it J.~Chem.~Phys.}~{\bf #1},\ #2\ (#3)}

\def\pra#1#2#3{{\it Phys.~Rev.}~A~{\bf #1},\ #2\ (#3)}

\def\prl#1#2#3{{\it Phys.~Rev.~Lett.}~{\bf #1},\ #2\ (#3)}

\def\rmp#1#2#3{{\it Rev.~Mod.~Phys.}~{\bf #1},\ #2\ (#3)}

\def\bfrm#1{\textbf{\textrm #1}}

\def\colvecnext#1{
        #1
        \global\advance\colveccount-1
        \ifnum\colveccount>0
                \\
                \expandafter\colvecnext
        \else
                \end{pmatrix}
        \fi
}

\begin{document}

\title{
Gaussian Process Model for Collision Dynamics of Complex Molecules}
\author{Jie Cui}
\affiliation{Department of Chemistry, University of British Columbia, Vancouver, B.C., V6T 1Z1, Canada}

\author{Roman V. Krems}
\affiliation{Department of Chemistry, University of British Columbia, Vancouver, B.C., V6T 1Z1, Canada}
\date{\today}

\begin{abstract}

We show that a Gaussian Process model can be combined with a small number (of order 100) of scattering calculations to provide a multi-dimensional dependence of scattering observables on the experimentally controllable parameters (such as the collision energy or temperature) as well as the potential energy surface (PES) parameters. { For the case of Ar - C$_6$H$_6$ collisions, we show that 200 classical trajectory calculations are sufficient to provide a 10-dimensional hypersurface, giving the dependence of the collision lifetimes on the collision energy, internal temperature and 8 PES parameters.} This can be used for solving the inverse scattering problem, the efficient calculation of thermally averaged observables, for reducing the error of the molecular dynamics calculations by averaging over the PES variations, and the analysis of the sensitivity of the observables to individual parameters determining the PES.
Trained by a combination of classical and quantum  calculations, the model  provides an accurate description of the quantum scattering cross sections, even near scattering resonances. 
%In this case, the classical calculations stabilize the model against uncertainties arising from wildly varying correlations of resonantly enhanced quantum results.

\end{abstract}

\maketitle

The reliable scattering calculations of dynamical properties of molecules are required in almost any research field related to molecular physics. In particular, the experiments on collisional cooling of molecules to cold and ultracold temperatures \cite{cold-molecules}, chemical reaction dynamics \cite{resonances-1}, the development of new pressure standards \cite{kirk-1,kirk-2}, astrophysics and astrochemistry \cite{astrophysics} rely on accurate calculations of molecular collision cross sections. Currently, there are two major problems with the ab initio calculations of molecular dynamics observables. The first problem is the inaccuracy of the potential energy surfaces (PES). Unfortunately, even the most sophisticated quantum chemistry calculations produce the PES with uncertainties that lead to significant (and often unknown) errors in the dynamical calculations. This sensitivity to PES inaccuracies is especially detrimental for low temperature applications (cold molecules, ultracold chemistry, astrophysics and pressure standards) \cite{hutson-pes,hutson-pes2,nh-nh}. The second problem is related to the numerical complexity of the quantum dynamics calculations \cite{ct,mott-massey}. For complex molecules with many degrees of freedom, accurate dynamical calculations are extremely time-consuming and it is often impossible to compute enough results for accurate averaging over the collision or internal energies of the colliding partners. 

In the present work we propose a solution to these two problems. In order to account for the PES uncertainties, the dynamical results can be averaged over variations of the PES. If the computed observables are averaged over variations of each individual PES parameter, producing an {\it expectation interval} of the observables, the ab initio dynamical calculations can have fully predictive power (with error bars). 
However, the outcome of a molecular collision is generally a complicated function of many (ten or more) PES parameters. It is impossible to obtain the dependence of the collision observables on the individual PES parameters by the direct scattering calculations. We show that such a dependence can be obtained by a combination of a small number (on the order of 100) of scattering calculations with a Gaussian Process (GP) model \cite{sacks1989,rasmussen2004gaussian}. We show that the same model can be used to obtain the accurate dependence of the scattering observables on the collision or internal energies of the molecules, with a small number of scattering calculations. The result is an accurate global dependence of the scattering observables on the collision energy, internal energy and every individual parameter of the PES surface. This global dependence can be used to average the computed observables over variations of the {\it individual} PES parameters, as well as over the collision and internal energies in order to produce thermally average observables. It can also be used to analyze the influence of the individual PES parameters on the scattering outcome.  This makes the model proposed here a unique tool for the analysis of the effects of the PES topology on the molecular scattering dynamics.

Widely used in engineering technologies \cite{gp-applications-1,gp-applications-2}, the GP model can be viewed as a technique for interpolation in a multi-dimensional space. We choose the GP model because it is an efficient non-parametric method. There is no need to fit data by analytical functions so the model is expected to work for any distribution of scattering observables and to become more accurate when trained by more computed observables. 
Given the scattering observables computed at a small number of randomly chosen points in the multi-dimensional parameter space, the GP model learns from correlations between the values of these scattering observables to produce a smooth dependence on all the underlying parameters. 
As an illustrative example, we consider the scattering of benzene molecules C$_6$H$_6$ by rare gas (Rg) atoms He - Xe. The PES surface for C$_6$H$_6$ - Rg interactions is characterized by 8 parameters. We consider two scattering observables \cite{buffer-gas-cooling, bgc-2,bgc-3}: the collision lifetimes and the scattering cross sections. We address the following questions: how many scattering calculations are sufficient to train a GP model to produce an accurate global dependence on all the underlying parameters? Can the GP model be used to make predictions of the scattering observables for one collision system based on the known properties of another collision system? Can the GP model be used to characterize the scattering observables near quantum resonances?

%\section{GP model of scattering observables}

We consider a scattering observable ${\cal O}$ as a function of $q$ parameters described by vector $\bm x$. The components of the vector $\bm x = \left ( x_1, x_2,\cdots , x_q \right)^{\top}$ can be the collision energy, the internal energy and/or the parameters representing the PES. We assume that $\cal O$ is known from a classical or quantum dynamics computation at a small number of $\bm x$ values. Our first goal is to construct an efficient model that, given a finite set of ${\cal O}(\bm x)$, produces a global dependence of the scattering observable on $\bm x$. If the observable is known from a measurement or a rigorous quantum calculation as a function of some parameters $x_i$ -- {\it e.g.}, the collision energy -- we show that the model can be adjusted to produce the global dependence of ${\cal O}$ on $\bm x$ that reproduces the accurate data, even if the dynamical calculation method is inaccurate. 

We assume that the scattering observable of interest at any $\bm x$ is a realization of a Gaussian process $F(\cdot)$, characterized by a mean function $\mu(\cdot)$, constant variance $\sigma^2$ and correlation function $R(\cdot, \cdot)$. 
For any fixed $\bm x$, $F(\bm{x})$ is a value of a function randomly drawn from a family of functions Gaussian-distributed around $\mu(\cdot)$. Consequently, 
the multiple outputs $F(\bm{x})$ and $F(\bm{x}')$ at $\bm x$ and $\bm x'$ jointly follow a multivariate normal distribution defined by $\mu(\cdot)$,  $\sigma^2$, and $R(\cdot,\cdot)$ \cite{adler1981geometry, cramer2013stationary}. 
%Note that the variance $\sigma^2$ is the same for the outputs at $\bm x$ and $\bm x'$. 
%Since the correlation function is fitted to given data, the choice of $R(\cdot,\cdot)$ is somewhat flexible.
We assume the following form for the correlation function  \cite{mitchell1990existence, cressie1993statistics, stein1999interpolation,abt1999estimating}:
\begin{eqnarray}
R(\bm{x},\bm{x}') = \mathrm{exp} \left\{ -\sum_{i=1}^{q} \omega_{i}|x_i-x_i'|^{p} \right\}. 
\end{eqnarray}
and write 
\begin{eqnarray}
F(\bm{x}) = \sum_{j=1}^{k} {h}_{j}(\bm{x})\beta_{j} + {Z}(\bm{x}) = {\bf{h}}(\bm{x})^\top{\boldsymbol{\beta}} + {Z}(\bm{x}),
\label{GP-F}
\end{eqnarray}
where ${\bf h} = \left ( h_1(\bm x), ..., h_k(\bm x) \right )^\top$ is a vector of $k$ regression functions \cite{regressors}, 
$\boldsymbol{\beta} = (\beta_1,\beta_2,\cdots, \beta_k)^\top$ is a vector of unknown coefficients, and $Z(\cdot)$ is a Gaussian random function with zero mean.
The problem is thus reduced to finding $\boldsymbol{\beta}$, $p$ and $\boldsymbol{\Omega}=(\omega_1,\omega_2,\cdots,\omega_q)^\top$. 

We spread $n$ input vectors $\bm x_1, ..., \bm x_n$ evenly throughout a region of interest and compute the desired observable $\cal O$ at each $\bm x_i$ with a classical or quantum dynamics method. The outputs of a GP at these points  $\bm{Y}^n=\Big({F}(\bm{x}_1), 
{F}(\bm{x}_2),\cdots,{F}(\bm{x}_n)\Big)^{\top}$ follow a multivariate normal distribution
%\begin{eqnarray}
%\bm{Y}^n \thicksim \mathrm{MVN} (\mathbf{H}\boldsymbol{\beta},\sigma^2 \mathbf{A})
%\end{eqnarray}
with the mean vector $\mathbf{H}\boldsymbol{\beta}$ and the covariance matrix 
$\sigma^2 \mathbf{A}$. Here, $\mathbf{H}$ is a $n\times k$ design matrix with $i$th row filled with the $k$ regressors $h_1(\bm{x}_i),h_2(\bm{x}_i),\cdots,h_k(\bm{x}_i)$ at site $\bm{x}_i$ , 
and $\mathbf{A}$ is a $n\times n$ matrix with the elements ${\bf A}(i,j) = R(\bm x_i, \bm x_j)$.

%\begin{eqnarray}
%\mathbf{A} =  \left(\begin{array}{cccc} 1 & R(\bm x_1, \bm x_2) & \cdots & R(\bm x_1, \bm x_n) \\  R(\bm x_2, \bm x_1) & 1& \ & \vdots \\ \vdots&  & \ddots &  \\R(\bm x_n, \bm x_1) & \cdots &  & 1 \\\end{array} \right)
%\end{eqnarray}

Given $\boldsymbol{\Omega}$, the maximum likelihood estimators (MLE) of $\boldsymbol{\beta}$ and $\sigma^2$ have closed-form solutions \cite{sacks1989}:
\begin{eqnarray}
\boldsymbol{\hat{\beta}} (\boldsymbol{\Omega}) = (\mathbf{H}^{\top}\textbf{A}^{-1}\mathbf{H})^{-1}\mathbf{H}^{\top}\textbf{A}^{-1}\textbf{\textit{Y}}^n
\end{eqnarray}
\begin{eqnarray}
\hat{\sigma}^2 (\boldsymbol{\Omega}) = \frac{1}{n}(\textbf{\textit{Y}}^n-\mathbf{H}\boldsymbol{\beta})^{\top}{\bf{A}}^{-1}(\textbf{\textit{Y}}^n-\mathbf{H}\boldsymbol{\beta})
\end{eqnarray} 
To find the MLE of $\boldsymbol{\Omega}$, we fix $p$ and maximize the log-likelihood function
\begin{eqnarray}
\textrm{log}\mathcal{L}(\boldsymbol{\Omega}|\bm{Y}^n)=-\frac{1}{2}\left[n\textrm{log}\hat{\sigma}^2+ \textrm{log}(\textrm{det}(\textbf{A})) + n \right]
\end{eqnarray}
numerically by an iterative computation of the determinant  $|\textbf{A}|$ and the matrix inverse $\textbf{A}^{-1}$.

The goal is to make a prediction of the scattering observable at an arbitrary $\bm x = \bm x_0$. 
Because the values $Y_0=F(\bm{x}_0)$ at $\bm x_0$
and the outputs at training sites 
%$\bm{Y}^n=\Big({F}(\bm{x}_1), {F}(\bm{x}_2),\cdots,{F}(\bm{x}_n)\Big)^{\top}$ 
are jointly distributed, 
%as
%\begin{eqnarray}
%\mat{Y_0\\ \bm{Y}^n}  \thicksim \mathrm{MVN} \left(\mat{\textbf{\textrm{h}}(\bm{x}_0)^{\top} \\ \textbf{H}}\boldsymbol{\beta}
%, \sigma^2 \mat{1 & \mathbf{A}_0^{\top}\\ \mathbf{A}_0&\mathbf{A} }\right)
%\end{eqnarray}
%where 
 %$\mathbf{A}_0=(R(\bm{x}_0,\bm{x}_1),R(\bm{x}_0,\bm{x}_2),\cdots,R(\bm{x}_0,\bm{x}_n))^{\top}$ is specified by the now known correlation function $R(\cdot|\boldsymbol{\hat \Omega})$. 
%This means that 
the conditional distribution of possible values $Y_0=F(\bm{x}_0)$ given the values $\bm{Y}^n$ is a normal distribution 
%\begin{eqnarray}
%Y_0|\bm{Y}^n, \boldsymbol{\beta}, \sigma^2, \boldsymbol{\Omega}\thicksim \mathrm{N} (m(\bm{x}_0)^*,\sigma^{*2}_z(\bm{x}_0))
%\end{eqnarray}
with the conditional mean and variance
\begin{eqnarray}
\label{mean-prediction}
m(\bm{x}_0)^*&=& \bfrm{h}(\bm{x}_0)^{\top}\boldsymbol{\beta}+\mathbf{A}_0^{\top}\mathbf{A}^{-1}(\bm{Y}^n -\textbf{H}\boldsymbol{\beta} ) \\
\sigma^{*2}(\bm{x}_0) &=& \sigma^2 (1- \mathbf{A}_0^{\top}\mathbf{A}^{-1}\mathbf{A}_0),
\end{eqnarray}
where 
$\mathbf{A}_0=(R(\bm{x}_0,\bm{x}_1),R(\bm{x}_0,\bm{x}_2),\cdots,R(\bm{x}_0,\bm{x}_n))^{\top}$ is specified by the now known correlation function $R(\cdot|\boldsymbol{\hat \Omega})$. 
Eq. (\ref{mean-prediction}) provides the GP model prediction for the value of the scattering observable at $\bm x_0$

%{\it Model verification.}

To illustrate the applicability and accuracy of the GP model, we first compute the collision lifetimes of benzene molecules with Rg atoms \cite{zhiying, cui2012collisional,zrr}. We use the classical trajectory (CT) method described in Ref. \cite{zhiying}. As shown in Ref. \cite{potentials}, the C$_6$H$_6$ - Rg PES can be expressed as a sum over terms describing the interaction of Rg with the C-C and C-H bond fragments, characterized by $8$ parameters.  
We first fix the PES parameters to describe the C$_6$H$_6$ - Ar system and focus on the dependence of the lifetimes on two parameters: the collision energy $E$ and the rotational temperature $T_r$. Figure 1 shows the results of the CT calculations illustrating that the collision lifetime exhibits an inverse correlation with $E$, while no apparent correlation with $T_r$. 
Figure 1 (c) shows the global surface of the lifetime as a function of $E$ and $T_r$ obtained from the GP model with $h_1 = 1, h_{i > 1} = 0$ 
%reducing the first term in Eq. (\ref{GP-F}) to a constant $\beta$, 
and $p$ set to 1.95. To quantify the prediction accuracy of the GP model, we calculate the errors
%the empirical root mean squared error (ERMSE)
 $\varepsilon_E = \sqrt{\frac{1}{n}\sum_{i=1}^{n}(y_i-\hat{y}_i)^2}$
%and the scaled root mean squared error (SRMSE)
and $\varepsilon_S=\varepsilon_E/{(y_{\rm max}-y_{\rm min})}$, where $y_i$ are the computed values and $\hat{y}_i$ are the GP model predictions. 
For the model with only 20 scattering calculations used as training points, 
$\varepsilon_E = 9.36$ ps and $\varepsilon_S = 7.93~\%$. With the number of the scattering calculations increased to 50, the errors decrease to $\varepsilon_E = 5.17$ ps and $\varepsilon_S = 4.38~\%$.

\begin{figure}[ht]
\label{figure1}
\begin{center}
\includegraphics[scale=0.25]{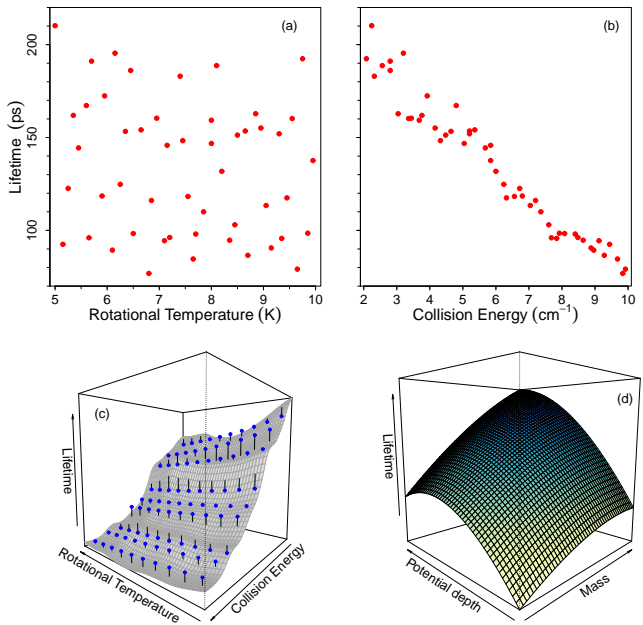}
\end{center}
\caption{
(a) and (b): 
The lifetime dependence on the rotational temperature  and collision energy for C$_6$H$_6$ -- Ar collisions.
(c): 
The surface produced by the GP model.  The lines connect the values (circles) computed from the classical trajectories with the values predicted by the GP model. 
(d): The surface produced by the GP model for 
C$_6$H$_6$ -- Rg collision lifetimes vs the atomic mass and the PES depth for $T_r = 4$ K  and $E = 4$ cm$^{-1}$. { The surface (c) is produced with only 20 scattering calculations on input and has the normalized error $ \varepsilon_S < 8$ \%. The surface (d) is produced with 40 scattering calculations and has the error $ \varepsilon_S = 5.09$ \%.} 
}
\end{figure}

%The data summarized in Table 2 show that the GP model with only 20 training sites produces a global surface giving the dependence of the collision lifetime on the rotational temperature and the collision energy accurate to within 10 \%.

%In addition, due to randomness, different samples gives different predicted outputs even for the sample size, which explains the slight increase in the prediction error when sample size increase from 20 to 50 for $p=2$.

%\section{Scattering data interpolation}

%Figure 1 illustrates the power of the GP model for the analysis of scattering observables. 
The scattering calculations presented in the upper panels of Figure 1 cannot be interpreted to assume any simple functional form. In addition, the vastly different gradients of the $T_r$ and $E$ dependence may make the conclusions based on calculations at fixed values of one of the parameters misleading. In contrast, the surface plot in  Figure 1(c) clearly illustrates that the collision lifetimes 
decrease monotonically with both $T_r$ and $E$. The effect of the rotational temperature is much weaker especially when $E > 5$ cm$^{-1}$ and there is no strong two-way interaction between $T_r$ and $E$.
The GP model surface can be used to evaluate thermally averaged collision lifetimes by integrating the $E$-dependence at given $T_r$.

The GP model can be extended to multiple collision systems for the predictions of the collision properties of a specific collision system based on the known collision properties of another system. To illustrate this, we consider the lifetimes of the long-lived complexes formed by benzene in collisions with Rg atoms He -- Xe. 
As the collision system is changed from C$_6$H$_6$ - He to C$_6$H$_6$ - Xe, there are two varying factors that determine the change of the collision dynamics: the reduced mass and the PES. 
%These factors must be correlated but the correlations cannot be clearly determined from the scattering calculations for different systems because the dependence of the scattering observable on these two parameters is very different. 

As before, we use the GP model $F(\bm x) = \beta + Z(\bm x)$, with $\bm x$ now representing the atomic mass $\mu_A$ and the interaction strength $D_e$ at the global minimum of the atom - molecule PES obtained by scaling the Ar - C$_6$H$_6$ PES. 
We fix $T_r = 4$ K  and $E = 4$ cm$^{-1}$, and compute the collision lifetimes 
at 40 randomly chosen points in the interval of  $\mu_A$ and $D_e$
$[4$g/mol$,130$g/mol$]\times[80\textrm{cm}^{-1},520\textrm{cm}^{-1}]$, which covers all of
the Rg -- C$_6$H$_6$ systems.  These 40 calculation points are then used to train the GP model to produce the surface plot shown in Figure 1 (d). The error $\varepsilon_S$ of the surface is $5.09$ \%. 
The plot reveals that increasing both $\mu_A$ and $D_e$ enhances the collision lifetimes and that 
the reduced-mass dependence of the collision lifetimes is very weak compared to the dependence on the interaction strength.

\begin{figure}[ht]
\label{figure2}
\begin{center}
\includegraphics[scale=0.25]{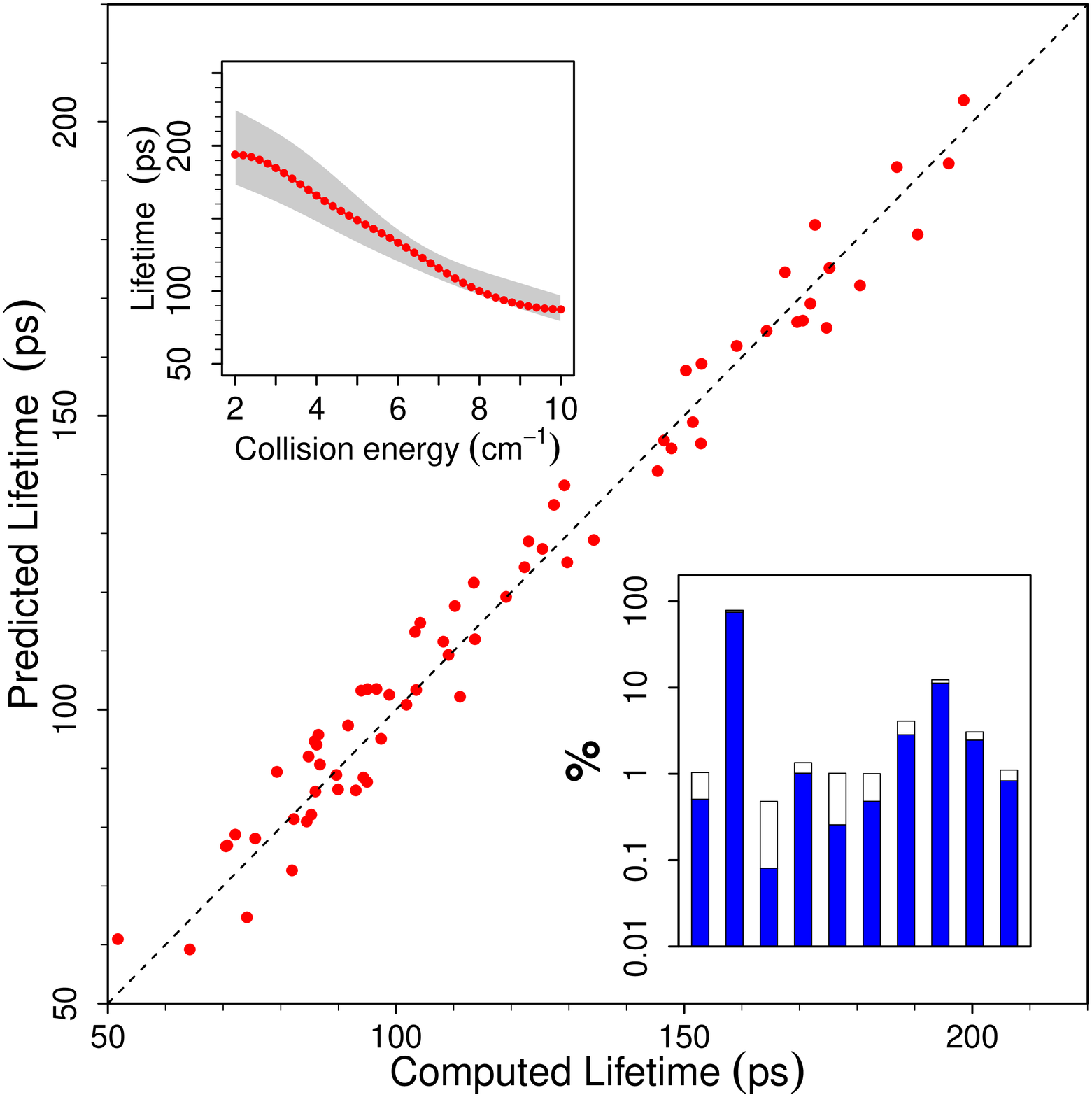}
%\vspace{-7.5cm} \hspace{-1.7cm} 
%\includegraphics[scale=0.15]{./Figures/figure2-inset-2} \\
%\vspace{+1.cm} \hspace{3.cm}
%\includegraphics[scale=0.15]{./Figures/figure2-inset}
%\vspace{0.4cm}
\end{center}
\caption{
Accuracy of the GP model with variable PES parameters for the prediction of
 the collision lifetimes. The scatter plot compares the predicted values with the computed values. 
The error of the GP model is the deviation of the points from the diagonal line. 
{ This GP model is trained by only 200 scattering calculations, enough to produce a 10-dimensional hypersurface with the error $\varepsilon_S = 4$ \%.}
Left inset: Energy dependence of the collision lifetime for Ar - C$_6$H$_6$ with the error interval obtained by varying {\it all} the individual PES parameters by $\pm 3~\%$.
Right inset: Relative effect of the variation of $T_r$, $E$ and the PES parameters on the collision lifetimes. The filled area of the bars shows the uncorrelated contribution of the corresponding variable and the open area --  the effect that depends on one or more other variables.
}
\end{figure}

%\section{Multi-dimensional model}

The GP model can be exploited to explore the role of the individual PES parameters on the observables. To illustrate this, we now consider that $\bm x$ contains $8$ parameters giving the analytical form of the Rg - C$_6$H$_6$ PES \cite{potentials}, in addition to $E$ and $T_r$. 
We calculate the lifetimes at 200 randomly selected points in this parameter space and use these points to train the GP model.  Figure 2 compares the predicted values with the calculated values for another set of 70 randomly selected points. The plot corresponds to the model error $\varepsilon_S = 4~\%$.

%\begin{figure}[ht]
%\label{figure3}
%\begin{center}
%\includegraphics[scale=0.3]{./Figures/figure3}
%\end{center}
%\caption{
%Relative effect of the variation of the rotational temperature, collision energy and the PES parameters ($t_1$ -- $t_8$) on the collision lifetimes. The blue area of graph shows the uncorrelated contribution of the corresponding variable and the red area -- the part of the effect that depends on one or more other variables.  
%}
%\end{figure}

The 10-parameter GP model contains a wealth of information on the dependence ${\cal O}(\bm x)$. For example, one can perform a sensitivity analysis by using the functional analysis of variance decomposition \cite{pujol2008sensitivity, saltelli2008global, roustant2012dicekriging}
to determine, which of the PES parameters have the strongest impact on the observable (right inset of Figure 2).  Of the 8 PES parameters, the location of the potential well due to the interactions of Rg with the C-C bonds for the parallel approach \cite{potentials} is the most important factor determining the collision lifetime. The model can also use be used to compute the uncertainties due to {\it global} variation of the PES. Figure 2 (left inset) shows the interval of the lifetimes obtained by the simultaneous $\pm 3~\%$ variation of all $8$ PES parameters.

%\section{Fitting experimental data}

{
We now consider the applicability of the GP model to quantum scattering calculations. The quantum results are often affected by resonances \cite{resonances-1,resonances-2}, leading to wild variations of the scattering observables in a small range of the underlying parameters. If applied directly to such the case, the GP model is unstable because steep variation of the correlations leads to singularities in ${\bf A}^{-1}$  \cite{hankin}. This is illustrated in Figure 3, showing the GP model predictions trained directly by 60 quantum calculations of cross sections for rotationally inelastic He - C$_6$H$_6$ scattering, randomly chosen at $E$ between 1 and 10 cm$^{-1}$.  
The instability of the GP model arises from the wild variations of the scattering cross sections near resonances. We repeated these calculations for the elastic and state-resolved rotationally inelastic cross sections shown in Figure 4 (a-c) of Ref. \cite{zrr}. In each case, we found that the wild variation of the quantum results leads to unstable GP model predictions. 
}

However, the GP model can be extended to model the time-consuming quantum scattering calculations with the help of efficient classical dynamics calculations. To do this, we introduce a more complex GP
 as \cite{kennedy2001bayesian}
\begin{eqnarray}
E(\cdot)  = \rho F(\cdot) + G(\cdot) + \varepsilon,
\label{GP-exp}
\end{eqnarray}
where $F(\cdot)$ and $G(\cdot)$ are independent Gaussian random functions, with $G(\cdot)$ characterizing the difference between the CT and QM  calculations and effectively describing the inaccuracy of the classical trajectory method. The calculations are performed in two steps. First, the CT calculations are used to train the GP model $F(\cdot)$. In the second step, 
the QM and CT calculations are used together to train the model $G(\cdot)$ in Eq. (\ref{GP-exp}), using the parameters of $F(\cdot)$ and treating $ \rho$ and $\varepsilon$ as variable parameters. This fixes the models $F(\cdot)$ and $G(\cdot)$ as well as  $ \rho$ and $\epsilon$. 

The accuracy of this combined quantum - classical model is illustrated in Figure 3, showing that the model provides an accurate energy dependence of the cross sections, even near scattering resonances. The CT calculations in a two-function model (\ref{GP-exp}) stabilize the model, removing the errors arising from the resonant variation of the quantum results. { We applied the two-step model (\ref{GP-exp}) to the calculations for the elastic and state-resolved rotationally inelastic cross sections shown in Figure 4 (a-c) of Ref. \cite{zrr} and found a similar improvement in each case. }

\begin{figure}[ht]
\label{figure3}
\begin{center}
\includegraphics[scale=0.35]{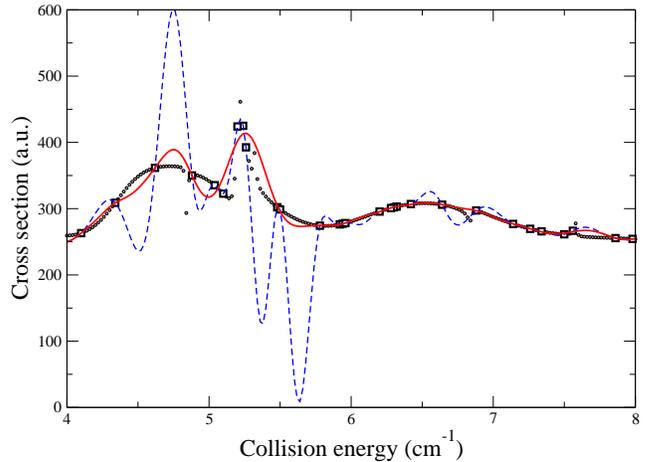} 
\end{center}
\caption{
{ GP models (solid curves) of quantum scattering cross sections (symbols) for C$_6$H$_6$ - He collisions. Blue dashed line: quantum calculations are used directly to train the GP model  (\ref{mean-prediction}). Red solid line: A combination of classical and  quantum results is used in a hybrid GP model (\ref{GP-exp}). The CT results stabilize the GP model predictions of the quantum calculations. The models are trained by the points represented by squares. The circles are used to illustrate the accuracy.}
}
\end{figure}

In summary, we have shown that a Gaussian Process model combined with a small number of scattering calculations can be used to obtain an accurate multi-dimensional dependence of the scattering observables on the experimentally controllable parameters and the PES parameters. 
{
Specifically, we showed that the GP model trained only by 20 CT calculations produces a dependence of the C$_6$H$_6$ - Ar collision lifetimes on the collision energy and the rotational temperature of benzene, with the normalized error $\varepsilon_S < 8~\%$. Trained by 200 calculations, the GP model produces a 10-dimensional dependence of the collision lifetimes on the collision energy, the rotational temperature and 8 individual  PES prameters, with the error $\varepsilon_S < 4$ \%. We have introduced a hybrid GP model that can be trained by  a combination of classical and quantum dynamics calculations in order to model the quantum results.} We showed that this model works even in the vicnity of quantum scattering resonances, where the direct fit of the quantum results by means of a GP model is unstable. 
%The GP model can thus be used to connect classical and quantum dynamics calculations, allowing one to use classical calculations to interpolate quantum results. 
The models described here 
are expected to find a wide range of applications, from fitting the interaction potentials by solving the inverse scattering problem, to analyzing the dependence of scattering observations on external parameters, to calibrating the accuracy of the scattering calculation methods. For example, the inverse scattering problem can be approached with the help of Eq. (\ref{GP-exp}), where $F(\cdot)$ is parametrized by unknown PES parameters and $E(\cdot)$ models the experimental data. The best estimates of the unknown PES parameters can then be found by a Markov-chain Monte Carlo method \cite{MCMC}, in a procedure similar to one recently applied in Ref. \cite{hep}.

\begin{acknowledgments}

We thank Dr. Zhiying Li for allowing us to use her codes for the classical and quantum dynamics calculations. This work is supported by NSERC of Canada. 

\end{acknowledgments}


\begin{thebibliography}{99}
\bibitem{regressors}
The regression functions can be chosen to follow a specific parametric dependence of the calculated data in order to make the GP model more efficient. 
If no such dependence is known or can be determined, the regression functions can simply be chosen as $h_1 = 1$ and $h_{i > 1} = 0$, reducing the first term in Eq. (\ref{GP-F}) to a constant $\beta$.
\end{thebibliography}
\end{document}